\documentclass[twocolumn]{aastex62}

\usepackage{ae,aecompl}
\usepackage{epsfig}
\usepackage{amsmath}
\usepackage{graphicx}
\usepackage{url}
\usepackage{amssymb}
\usepackage{times}
\usepackage{float}
\usepackage{natbib}
\usepackage{multirow}
\usepackage{booktabs}

\makeatletter
\newcommand{\vo}{\vec{o}\@ifnextchar{^}{\,}{}}
\makeatother

\def\beq{\begin{equation}}
\def\eeq{\end{equation}}
\def\beqar{\begin{eqnarray}}
\def\eeqar{\end{eqnarray}}

\def\msol{M_\odot}

\def\isotope#1#2{\mbox{${}^{#2}{\rm #1}$}}
\def\fe5#1{\isotope{Fe}{5#1}}
\def\co5#1{\isotope{Co}{5#1}}
\def\ni5#1{\isotope{Ni}{5#1}}

\def\fun#1#2{\lower3.6pt\vbox{\baselineskip0pt\lineskip.9pt
 \ialign{$\mathsurround=0pt#1\hfil##\hfil$\crcr#2\crcr\sim\crcr}}}

\shorttitle{MeV Gamma Rays from Fission}
\shortauthors{Wang, Vassh, et al.}

\begin{document}

\title{MeV Gamma Rays from Fission:\\
A Distinct Signature of Actinide Production in Neutron Star Mergers}

\correspondingauthor{Xilu Wang, Nicole Vassh}
\email{xwang50@nd.edu, xlwang811@gmail.com, \\ nvassh@nd.edu, nvassh@gmail.com}

\author[0000-0002-5901-9879]{Xilu Wang}
\altaffiliation{N3AS Postdoctoral Fellow}
\affil{Department of Physics, University of California, Berkeley, CA 94720, USA}
\affil{Department of Physics, University of Notre Dame, Notre Dame, IN 46556, USA}
\collaboration{(N3AS collaboration)}

\author{Nicole Vassh}
\affil{Department of Physics, University of Notre Dame, Notre Dame, IN 46556, USA}
\collaboration{(FIRE collaboration)}

\author{Trevor Sprouse}
\altaffiliation{Los Alamos CSES student fellow}
\affil{Department of Physics, University of Notre Dame, Notre Dame, IN 46556, USA}
\affil{Theoretical Division, Los Alamos National Laboratory, Los Alamos, NM, 87545, USA}

\author{Matthew Mumpower}
\affiliation{Theoretical Division, Los Alamos National Laboratory, Los Alamos, NM, 87545, USA}
\affiliation{Center for Theoretical Astrophysics, Los Alamos National Laboratory, Los Alamos, NM, 87545, USA}

\author{Ramona Vogt}
\affil{Lawrence Livermore National Laboratory, 7000 East Avenue, Livermore, CA 94550, USA}

\author{Jorgen Randrup}
\affil{Lawrence Berkeley National Laboratory, Nuclear Science Division, 1 Cyclotron Road, Berkeley, CA 94720, USA}

\author{Rebecca Surman}
\affil{Department of Physics, University of Notre Dame, Notre Dame, IN 46556, USA}

\begin{abstract}
Neutron star mergers (NSMs) are the first verified sites of rapid neutron capture ($r$-process) nucleosynthesis, and could emit gamma rays from the radioactive isotopes synthesized in the neutron-rich ejecta.
These MeV gamma rays may provide a unique and direct probe of the NSM environment as well as insight into the nature of the $r$ process, just as observed gammas from the $^{56}$Ni radioactive decay chain provide a window into supernova nucleosynthesis.
In this work, we include the photons from fission processes for the first time in estimates of the MeV gamma-ray signal expected from an NSM event.
We consider NSM ejecta compositions with a range of neutron richness and find a dramatic difference in the predicted signal depending on whether or not fissioning nuclei are produced.
The difference is most striking at photon energies above $\sim3.5$ MeV and at a relatively late time, several days after the merger event, when the ejecta is optically thin. 
We estimate that a Galactic NSM could be detectable by a next generation gamma-ray detector such as AMEGO in the MeV range, up to $\sim10^4$ days after the merger, if fissioning nuclei are robustly produced in the event.
\end{abstract}


\section{Introduction} 
\label{sec:intro}

MeV gamma-rays ($E_{\gamma}\leq 10$ MeV) are emitted from newly synthesized radioactive nuclei and thus can provide a unique and direct probe of the nucleosynthesis and ejecta structure in astrophysical transients. One well-studied example of a transient nucleosynthetic event is a supernova \citep[e.g.,][and references therein]{Clayton1969, Bussard1989, The2014}. 
Supernovae, chiefly Type Ia, are dominant sources of iron-group elements \citep[e.g.,][and references therein]{Nomoto1984, Nomoto2013}, and thus the gamma rays mainly come from the radioactive decay chain of $^{56}$Ni to $^{56}$Co to $^{56}$Fe. 
Gamma-ray observations of supernovae can
precisely and directly measure $^{56}$Ni yields, and the light curves 
encode a map of the structure and mixing of the ejecta \citep[e.g.,][]{Bussard1989, The2014, Wang2019}.
Two nearby supernovae have been detected in MeV gamma rays so far,
core-collapse supernova SN1987A \citep{Matz1988, Teegarden1989} and
Type-Ia supernova SN2014J \citep{Churazov2014, Diehl2014}.

Similar to supernovae, gamma rays are also expected to be emitted from the isotopes synthesized in a neutron star merger (NSM) event. 
Unlike supernovae that produce mainly $^{56}$Ni, NSMs are expected to produce a broad range of heavy radioactive isotopes via rapid neutron capture ($r$-process) nucleosynthesis.
NSMs are the first verified $r$-process sites \citep{NSM}, after the multimessenger detection of GW170817, GRB 170817A and the electromagnetic counterpart SSS17a/AT2017gfo \citep{AbbottGW170817, Coulter2017, Goldstein2017, Savchenko2017, Valenti2017}. 
The optical and infrared observations of this event indicated lanthanide production \citep{Cowperthwaite2017, Kasen}; however, other than a tentative identification of strontium \citep{Watson2019}, no specific elemental yields could be determined \citep[e.g.,][]{Pian2017, Shappee2017}. Thus, while studies combining kilonova observations with other data from stars and simulations \citep[e.g.,][]{Thielemann2017}
make a strong case for NSMs to be a major or primary site of all $r$-process elements, there is still no direct evidence linking production of the heaviest elements such as gold, platinum, and the actinides with NSM outflows. Future observations may provide such evidence; \citet{Cfpaper} pointed out that if $^{254}$Cf is produced in the event, its spontaneous fission and subsequent fission product decays can dominate late-time nuclear reheating, leading to a longer and brighter near-infrared kilonova signal, as is also discussed in \citet{Wu1}. MeV gamma rays can provide complementary evidence, should the event occur nearby. Previous studies \citep{Hotokezaka2016, Li2019, Oleg1, Oleg2, Wu2} considered the MeV photons emitted from the $\beta$ decay of radioactive $r$-process species and found such signals can encode key information on composition and event morphology and are detectable out to $\sim$3-10 Mpc. 

An additional source of NSM MeV photons we consider here for the first time is the fission of actinide species. Fission reactions produce highly excited fragments that copiously emit neutrons and gammas over a large range of energies. If fissioning nuclei are present in the ejecta, the fission photons produced will contribute to the overall electromagnetic signal from the merger. This contribution is likely subdominant at early times, given the small number of fissioning species compared to those that $\beta$ decay. 
However, at later times ($\gtrsim$10 days), when the ejecta becomes optically thin,\footnote{The observations of SSS17a/AT2017gfo and theoretical studies \citep[e.g.,][]{Drout2017, Kilpatrick2017, Pian2017, Shappee2017, Waxman2019} suggest that the timescale when the ejecta becomes optically thin is about a few days after the merger moment.} and 
most $\beta$ decays with high $Q$ values, $Q_{\beta}$, are complete,
fission may become the dominant source for gamma-ray photons of multiple-MeV energies. Such a signal thus holds the potential to provide unambiguous evidence of actinide production in the merger event.

In this Letter we provide a first estimate of the late-time, high-energy MeV signal from a Galactic NSM event. We start from calculations of element synthesis and prompt spectra of photons emitted from fission and $\beta$ decay. We process the emitted spectra through a semianalytical radiation transfer model to estimate the time at which the MeV photons free-stream and simulate potential light curves and spectra. 
Finally we evaluate the detectability of a fission gamma signal in the next generation MeV telescope AMEGO \citep{AMEGO}\footnote{\url{https://asd.gsfc.nasa.gov/amego/index.html}} and find that a Galactic event would be detectable in the MeV range up to $\sim10^4$ days after the merger.

\section{Prompt Gamma-Ray Spectra of\\ \lowercase{$r$}-process nuclei}
\label{sec:spectra}

The energy release from fission is larger than other processes occurring in the $r$ process given
$Q\sim$200 MeV for fission. Although most of the energy released is allocated to the total kinetic energy (TKE) of the newly formed fission fragments ($\sim$ 170 MeV), these fragments are also created with a substantial excitation energy such that their de-excitation can easily lead to the prompt emission of neutrons and gammas in the MeV range. 
Fission gamma spectra exhibit a wide range of MeV gamma emission as the fragments de-excite toward their more stable ground-state configurations. High-energy statistical gammas are emitted through E1 (electric dipole radiation) and the giant dipole resonance while emission at energies less than $\sim$2-3 MeV are dominated by E2 (electric quadrupole radiation) transitions between nuclear levels.

\begin{figure}
\centering
\includegraphics[scale=0.55]{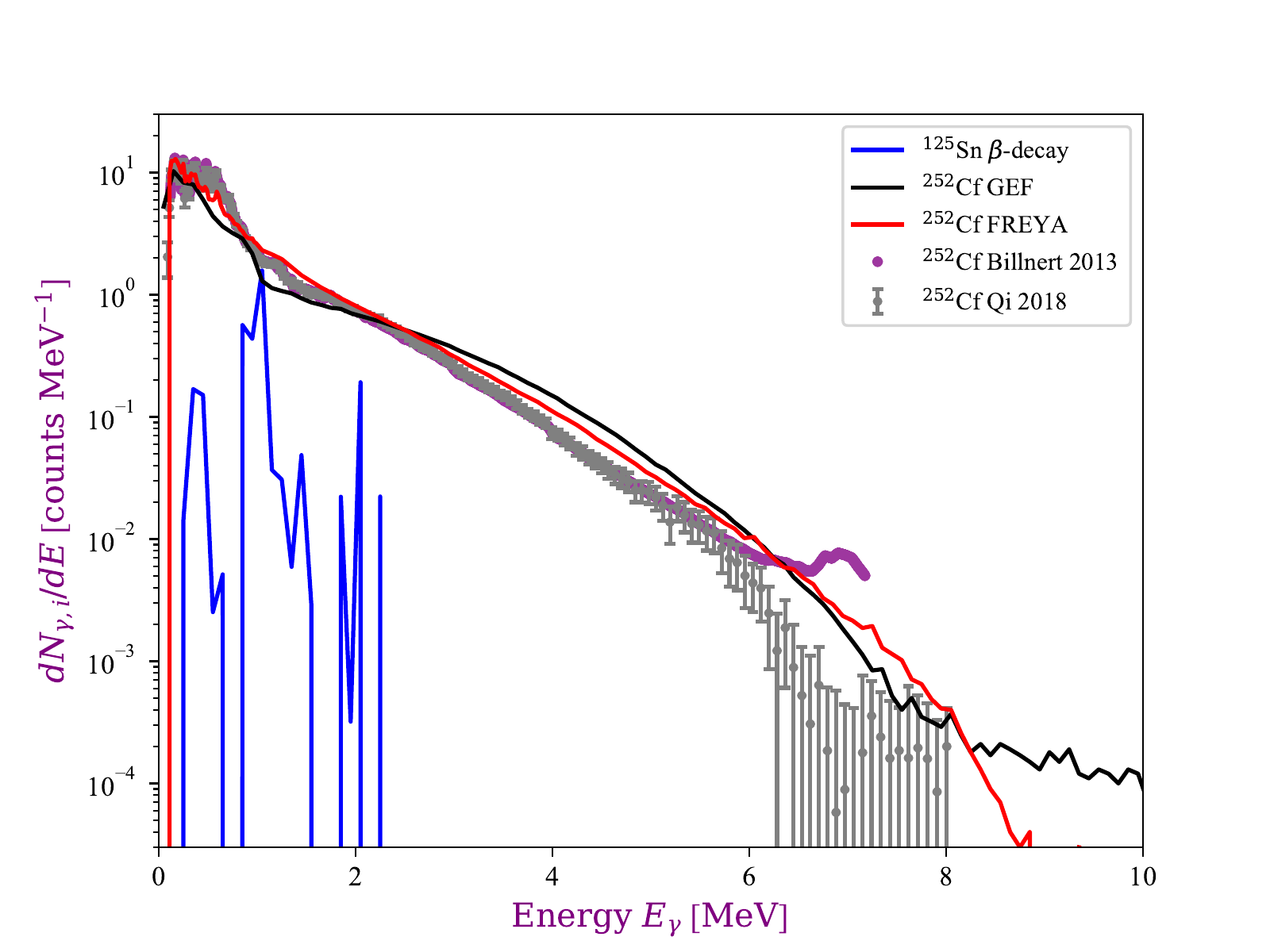}
\caption{The calculated prompt gamma spectra emitted from the spontaneous fission of $^{252}$Cf using GEF (black line) and FREYA (red line), as compared with experimental data from \cite{Billnert} (purple dots) and \cite{Qi} (grey dots). The prompt gamma spectra emitted from the $\beta$-decay of the $^{125}$Sn ground state (blue line) taken from ENDF/B-VIII.0 is also shown.
\label{fig:FREYA}
}
\end{figure}

An example fission gamma spectrum is shown in Figure~\ref{fig:FREYA} for $^{252}$Cf, a species that primarily decays through $\alpha$-decay but has a $3\%$ spontaneous fission branching ratio and a half-life of 2.6 yr. The prompt fission gamma spectrum of $^{252}$Cf is well-studied; two experimental data sets \citep{Billnert, Qi} are included in Figure~\ref{fig:FREYA}. As this example illustrates, fission can result in gamma emission with energies as high as $\sim$8 MeV.

Other decay processes that take place on the timescales of interest for detection of electromagnetic signals from NSMs are $\beta$ and $\alpha$ decay. In an $\alpha$ decay, most of the energy is released as the kinetic energy of the $\alpha$ particle and the decay radiation is dominated by sub-MeV X-rays. Thus here we consider how the gamma emission from fission is distinguishable from that of $\beta$ decay. 
We calculate the $\beta$-decay gamma-ray spectrum for each nucleus $i$ as
\beqar
\label{eq:beta}
\frac{dN_{\gamma,i}^{\beta^{-}}}{dE} & = & RP_{i,\beta^-}(E)
+ \sum\limits_{E'} \mathcal{I}_{i,\beta^-}(E')\delta(E-E'),
\label{eq:spectra1}
\eeqar
using a combination of individual lines at discrete energies $E$ and absolute intensity $\mathcal{I}(E)$ with a Dirac $\delta$ function distribution, as well as a possible continuum component $RP(E)$. 
In order to evaluate Eq.~\ref{eq:beta}, we require the spectra associated with individual decays encoded in the functions $\mathcal{I}$ and $RP$. For these $\beta$-decay spectra, we adopt data from  ENDF/B-VIII.0\footnote{\url{https://www-nds.iaea.org/public/download-endf/ENDF-B-VIII.0/}} \citep{ENDFB} and, when unavailable, we supplement with spectra calculated in the Los Alamos QRPA+HF framework \citep{Mumpower+16, MollerQRPA}. 

The distinct nature of the gamma emission from fission as compared to $\beta$-decay is evident from considering the $\beta$-decay gamma spectrum of a nucleus populated on timescales of days in the $r$ process, $^{125}$Sn. Exotic $r$-process nuclei with high $Q_{\beta}$ values typically decay on very fast ($\sim$seconds or less) timescales. As can be seen from Figure~\ref{fig:FREYA}, $\beta$-decay gamma spectra of interest at later times typically fall off after $\sim$3 MeV.
This presents an opportunity for gamma-ray telescopes to confirm the synthesis of heavy, fissioning actinides in a merger event through the detection of gammas in the $\sim$3-10 MeV energy range unique to the fission process.

The $r$ process potentially involves hundreds of neutron-rich actinides, and experimental data such as shown in Figure.\ref{fig:FREYA} are available for only very few relevant species. Therefore to obtain the individual fission gamma spectra for all the neutron-rich actinides that could be accessed by a merger event, we rely exclusively on theory and use the 2016 version of the code GEF \citep{GEF} as in \cite{VasshGEF2019}.
Since the energetics of prompt emission from excited fragments that determines the fission gamma spectra is an active area of study, we investigate model dependence using the FREYA code \citep{RVJR_gamma2,FREYA2}. We implement FREYA as described in \cite{VasshGEF2019} with the TKE and fission yields predicted by GEF taken as inputs for FREYA. The theoretical fission gamma spectra for $^{252}$Cf calculated with GEF and FREYA are compared to the experimental data in Figure~\ref{fig:FREYA}.

FREYA and GEF take different approaches to modeling the de-excitation treatment of the fission fragments. FREYA emits neutrons until energetically forbidden, given the one neutron separation energy, and then emits gamma-ray photons to dissipate the remaining excitation energy. GEF, instead, uses phenomenological decay widths of neutron and photon emission, allowing competition between the two. 
Additionally, the two codes differ in their treatments of level transitions that affect the low-energy fission gamma spectrum below $\sim$2 MeV. This region of larger differences between the treatments in the two codes, as seen in Fig.~\ref{fig:FREYA}, will be dominated by gammas emitted by $\beta$-decays in the $r$ process. From $\sim$3-8 MeV, where the GEF and FREYA results are similar, the spectrum is dominated by the giant dipole resonance. For energies above $\sim$8 MeV, the neutron-gamma competition in GEF leads to a pronounced high-energy tail relative to FREYA.
Importantly, despite all of the distinct approaches for modeling the de-excitation of fission fragments, the GEF and FREYA results for fission spectra are compatible from $\sim$3-8 MeV in showing a pronounced higher energy distribution. 
The consistency of the theoretical calculations and experimental data support our exploration of fission gamma emission at energies above 3 MeV as a potential $r$-process observable.

\section{Radiation Transfer Calculation}\label{sec:radiationTransfer}

\begin{figure*}
\centering
\includegraphics[width=1.25\columnwidth]{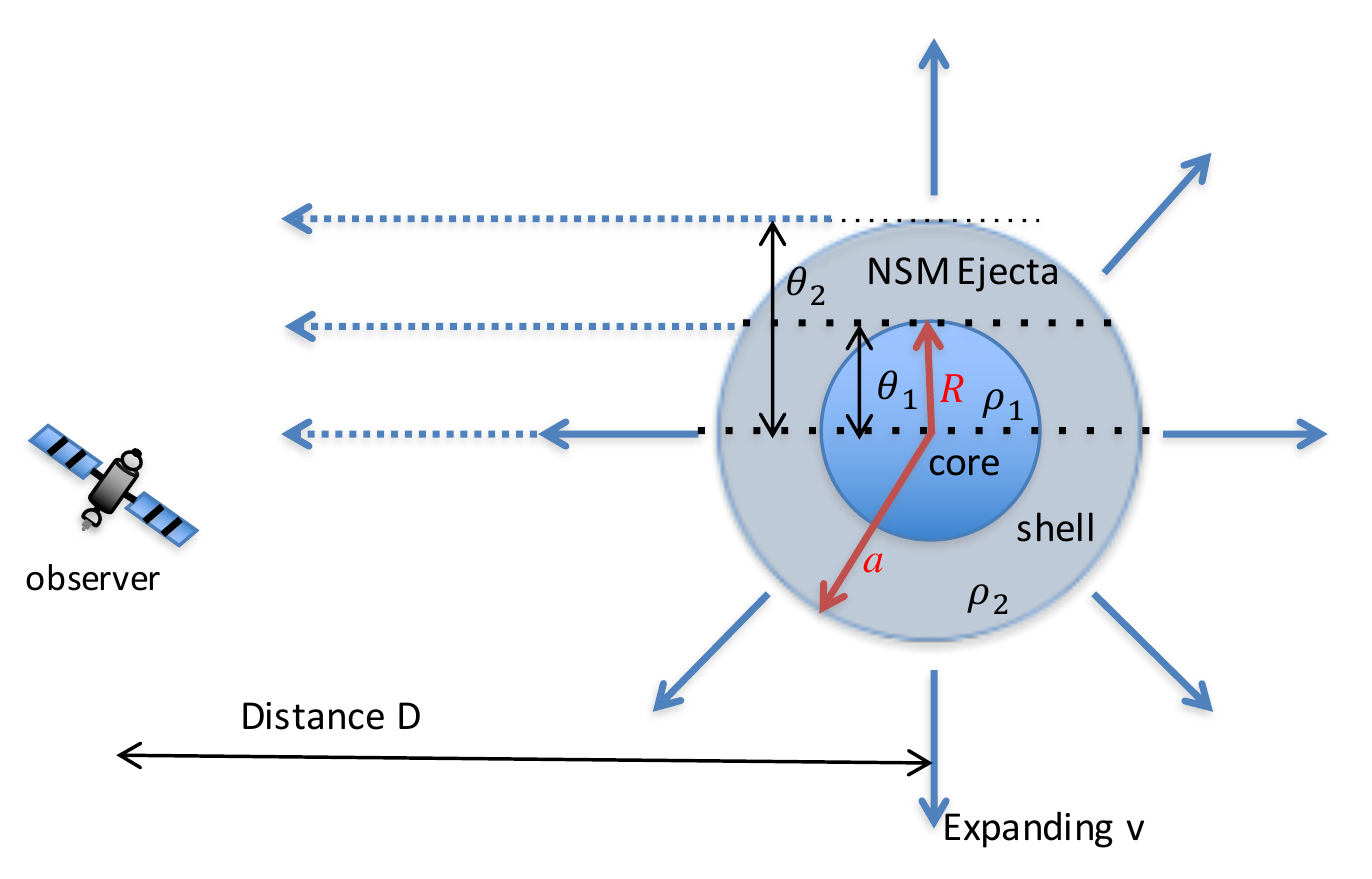}
\caption{
Sketch of the $r$-process ejecta model from an NSM. The distance between the observer and the NSM is $D$. In this model, the ejecta radius is $a$, subtended angle $\theta_2$. The $r$-process material is distributed in the shell of the ejecta with mass percentage $f$ and density $\rho_2$, and in the core with remaining mass, with radius $R$, subtended angle $\theta_1$, and density $\rho_2$. 
\label{fig:shell}
}
\end{figure*}

Since our aim is to investigate the role of fission in the gamma-ray emissions from an NSM, we focus on late times after $\sim10$ days when the gamma-ray signals due to fission could emerge.
Thus, a semianalytical calculation is adequate to compute the gamma-ray emissions observed after the propagation inside the NSM ejecta at this late time when the ejecta is nearly optically thin.

To estimate the light curve and spectrum observed on Earth, we adapt the radiation transfer calculations from \citet{Wang2019} with a uniform spherical ejecta model for the $r$-process ejecta.  
First, we assume the ejecta consists of an inner region with a higher density (core) plus an outer region with a lower density (shell), and 
is expanding homologously, i.e., $v(r)\propto r$. Here, the radius from the center is $r\sim vt$, and the maximum velocity at the outermost radius is $v_{0}$. 
The total ejecta mass is $M_{\rm ej}$, with a fraction $f$ of the mass distributed at the shell of the ejecta with radius $a(t)$ and density $\rho_{2}(t)$, while the remaining is in the core with radius $R(t)$, density $\rho_{1}(t)$, and subtended angle $\theta_1\sim R/D$, where $D$ is the distance between the Earth and the NSM. The angle subtended by the shell is between $\theta_1$ and $\theta_2\sim a/D$. Thus the ejecta emission is in the region between $\pm\theta_2$. A sketch of this ejecta model appears in Fig.~\ref{fig:shell}. 
These ejecta parameters will serve as inputs to the radiation transfer calculations, yielding estimates of the observable $\gamma$-ray light curves and spectra. 

Combining the prompt emitted photon spectra $dN_{\gamma,i}(E)/dE$ from each nucleus $i$ through fission and $\beta$ decay, with their relative nuclear abundances $Y_i(t)$ as functions of time $t$ and decay rates obtained from our $r$-process nucleosynthesis calculations, the emitted gamma-ray luminosity is then calculated as
\beqar
\label{eq:emitted}
L_{\gamma, \rm em}(E)&=&\frac{dN_{\gamma,\rm em}(E)}{dEdt}
\nonumber\\
&=&\frac{M_{\rm ej}}{m_{\rm p}} \Big(\sum_{i} \lambda_{i,\beta^-}Y_i(t) \times \frac{dN_{\gamma,i}^{\beta^{-}}(E)}{dE}
 \nonumber\\
& +&\sum_{j} \lambda_{j,\rm fission}{Y}_j(t) \times \frac{dN_{\gamma,j}^{\rm fission}(E)}{dE} \Big),
\eeqar
where $m_{\rm p}$ is the proton mass, $\lambda_{\beta^{-}}$ is the partial decay rate that is derived from the half-life and $\beta$-decay branching ratio, and $\lambda_{\rm fission}$ is the fission rate.
The nucleosynthetic yields are obtained using the nuclear reaction network code Portable Routines for Integrated nucleoSynthesis Modeling, or PRISM \citep{BDFrp}. Experimental half lives and rates are used whenever possible; otherwise we adopt the theoretical nuclear data applied in \citet{VasshGEF2019} and \citet{Wang2020} that makes use of GEF fission yields along with FRDM2012 masses and FRLDM fission barriers.

To propagate the emitted gamma rays through the ejecta, we solve the radiative transfer equation
$dI_{E}/dl=-\alpha(E) I_{E}+j_{E}$, 
with the absorption coefficient $\alpha(E) =\rho \kappa(E)$, where $\kappa(E)$ is the opacity of the dense ejecta material through which the photons propagate. Here, the source term is $j_{E}={L_{\gamma, \rm em}(E)}/{4\pi V}$.
The opacity values of the $r$-process isotopes are adopted from the XCOM website,\footnote{\scriptsize{\url{https://www.nist.gov/pml/xcom-photon-cross-sections-database}}}, for photon interactions including coherent (Rayleigh) scattering, incoherent (Compton) scattering, photoelectric absorption, and pair production. Only nonscattered photons are included in the observed gamma-ray signal here; scattered photons are ignored as their effects are minimal at late times when the ejecta is nearly optically thin.

We calculate the opacities in the MeV energy range of the $r$-process ejecta based on its composition with a mixture of the opacities of five characteristic isotopes (Fe, Xe, Eu, Pt and U).
That is, we adopt the opacity of iron for elements with mass number $A\leq109$, xenon for $110\leq A\leq137$, europium for $138\leq A\leq178$, platinum for $179\leq A \leq219$, and uranium for $A\geq220$.

The intensity $I$ along a line of sight with angle $\theta$, in the region with density $\rho$ and path length between $l_0$ and $l_m$ (the maximum path length for the photons in the line of sight is $l_{\rm max}=2\sqrt{a^2-(D\theta)^2}$), observed at energy $E_i$ after the Doppler shift is thus:
\beqar
\label{eq:intensity1}
I_{E_i}(\theta, l_m)=&&\frac{dN_{\gamma,\rm ob}}{dEdtdAd\Omega}
\nonumber\\
=&&I_{E_i}(\theta, l_0)e^{-(\tau(E_i, l_{\rm max})-\tau(E_i, l_0))}+
\nonumber\\
\int_{l_0}^{l_m}&&j_{E_0(l)} e^{-(\tau(E_0(l_m),l_{m})-\tau(E_0(l),l))}dl,
\eeqar
where the optical depth at the path length $l$ for the photon with energy $E$ is $\tau(E,l)=\int \rho_{\rm ej}\kappa(E) dl=\tau_a(E) l/a$, and $E_0$ is the energy of a photon that is emitted at the path length $l$ along the line of sight with angle $\theta$ and observed to be energy $E_i$, $E_0(\theta, l)={E_i}/[1+ ({v}/c) (l/a-{l_{\rm max}}/{2a})]$.

For the $r$-process ejecta sketched in Figure~\ref{fig:shell}, if the line of sight travels through the same region ($\theta_1\leq\theta\leq\theta_2$), the total intensity along the line of sight is $I_{E_i,1}(\theta)=I_{E_i}(\theta, l_{\rm max})$ with $l_0=0$. 
If the line of sight ($0\leq\theta\leq\theta_1$) travels through different regions (shell-core-shell), the total intensity $I_{E_i,2}(\theta)$ is calculated using Eq.~\ref{eq:intensity1} for each region in sequence, and the value of $I_{E_i}(\theta, l_m)$ calculated for the previous region serves as the initial condition $I_{E_i}(\theta, l_0)$ for the next region, from the path length $l=0$ until $l_{\rm max}$. 

Finally, the total flux from the $r$-process shell plus core ejecta is an integral of the intensity over the solid angle subtended by the ejecta, i.e., 
\beqar
\label{eq: flux}
F_{E_i}(t)&=& \frac{dN_{\gamma,\rm ob}}{dEdtdA}=\int I_{E_i}(\theta) cos\theta d\Omega 
\nonumber\\
&\approx& 2\pi \left(\int_{0}^{\theta_1} I_{E_i,1}(\theta) \theta d\theta +\int_{\theta_1}^{\theta_2} I_{E_i,2}(\theta) \theta d\theta \right),
\nonumber\\
F(t)&=&\int F_{E_i}(t)dE_i.
\eeqar

We select $M_{\rm ej}=0.01\msol$ \citep[e.g.,][]{Bovard}, $v_0=0.3c$ \citep[e.g.,][]{Shappee2017}, i.e., $\beta=0.3$, $\rho_{1,initial}=2\times10^{10} {\rm g/cm^3}$, and $\rho_{2,initial}=2\times10^{7} {\rm g/cm^3}$ \citep{PiranRoss, Rosswog} with shell fraction $f=10\%$. We use a fiducial distance $D=10$ kpc for a Galactic NSM; for other choices of distance, the flux and count rates will scale as $(10{\ \rm kpc}/D)^2$.
Variations in these astrophysical parameters result in only modest changes to our simulation results in terms of signal magnitude and the time when the eject becomes optically thin, and do not influence the qualitative conclusions of Section~\ref{sec:results}.

\section{Predicted MeV Gamma-Ray Signals from NSMs}
\label{sec:results}

\subsection{Very neutron-rich ejecta with robust fission}
\label{subsec:baseline}

\begin{figure}
\includegraphics[scale=0.55]{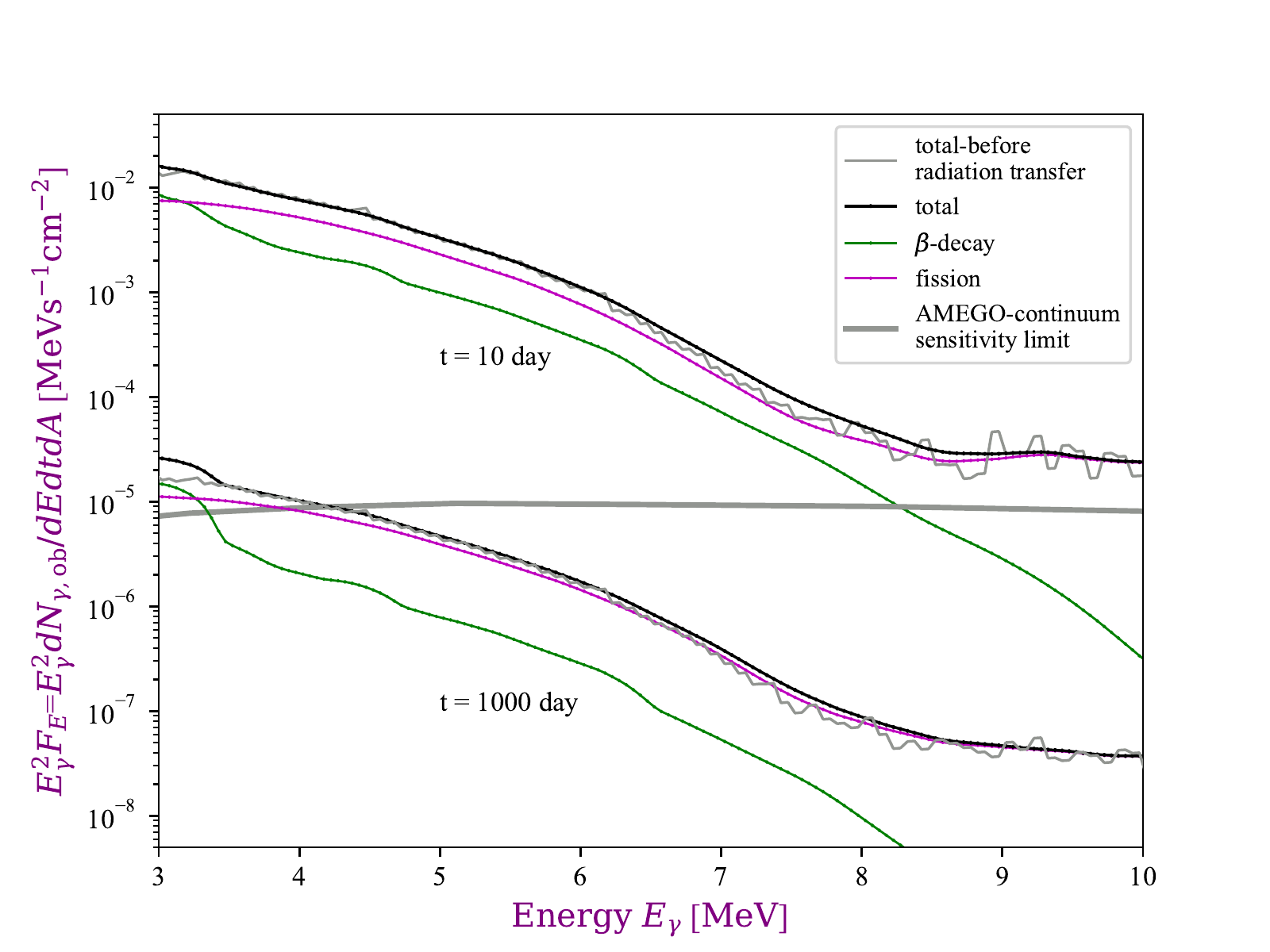}
\caption{Gamma-ray spectra at 10 and 1000 days given very neutron-rich ejecta from an NSM located at 10 kpc. Gray lines show the total prompt gamma-ray signal emitted, while black lines show the total signal after our radiation transfer calculation. Green lines show the contribution from $\beta$-decays to the observable gamma-ray signal, while the purple lines show the contribution from fission. The gray shaded band shows the sensitivity limit range for the AMEGO detector.
\label{fig:baselinespec}
}
\end{figure}

In very neutron-rich ejecta, the $r$ process pushes its reach into the heavy, neutron-rich actinides until the synthesis is ultimately terminated by fission.
Here we adopt conditions from the very neutron-rich ($Y_e\sim0.015$) dynamical ejecta tidal-tail simulations of \cite{Rosswog} and \cite{PiranRoss}. Our calculated low-energy ($\sim 0.15-3$) MeV spectra under these conditions are similar to published work \citep[e.g.,][]{Oleg1} and are dominated by low $Q_{\beta}$-value $\beta$-decays of $A<140$ nuclei. Figure~\ref{fig:baselinespec} shows the resulting spectra 
at higher energies of $>3$ MeV at 10 days and 1000 days after merger, calculated as described in Section~\ref{sec:radiationTransfer}. Comparing the contributions to the spectra from $\beta$-decay and fission, we find that above $\sim$3.5 MeV late-time gamma emission is dominated by high-energy fission gammas, as expected from Section~\ref{sec:spectra}. 
In addition, we find that the gamma-ray spectra from fission and $\beta$-decays are parallel between $\sim$3.5 and 8 MeV, demonstrating that at timescales on the order of days or longer the emission of $\beta$-decay photons in this energy range are a direct consequence of actinide production since they originate from the high $Q_{\beta}$-value decays of very neutron-rich, short-lived fission fragments.

\begin{figure}
\includegraphics[scale=0.55]{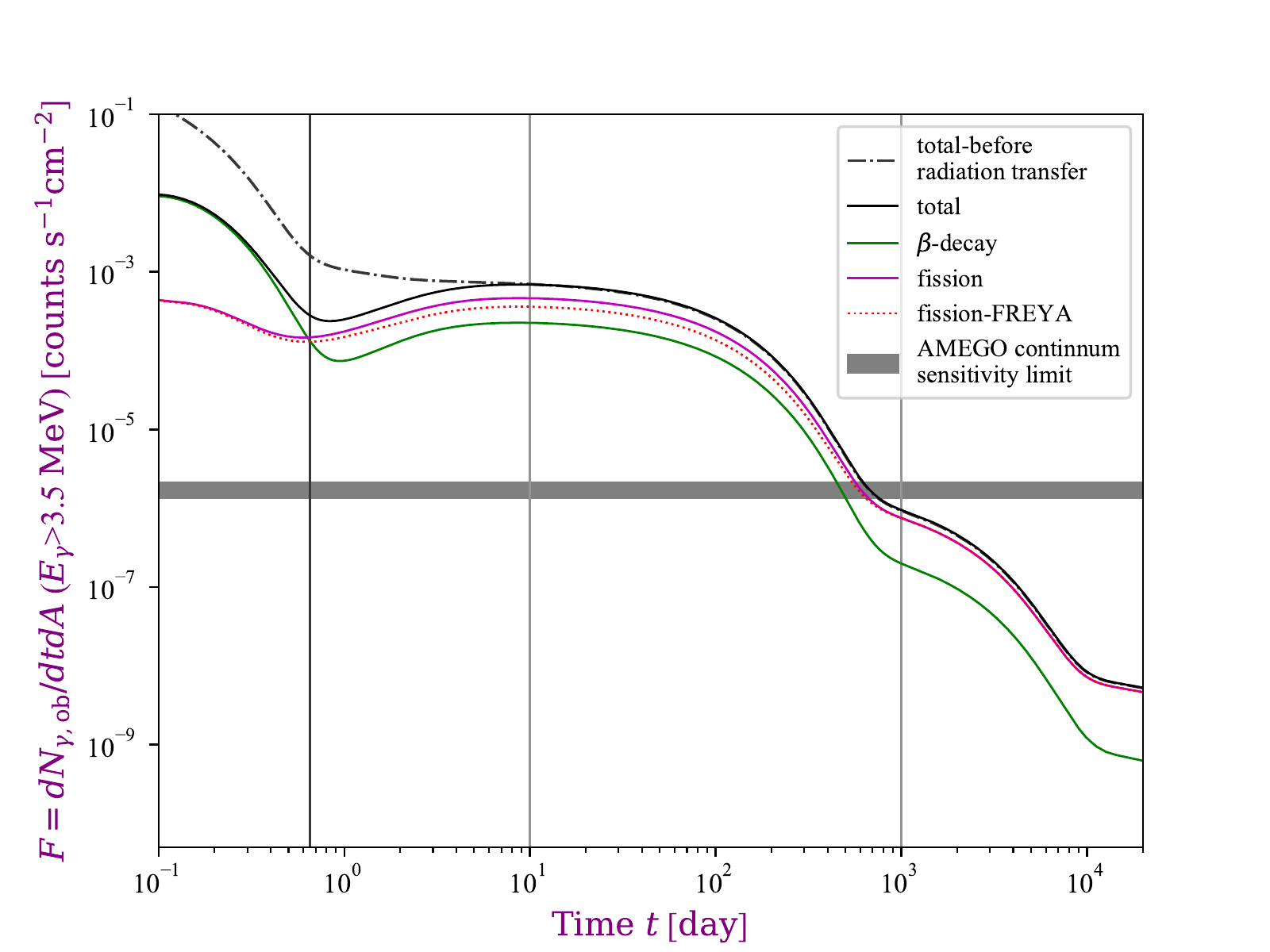}
\caption{The predicted gamma-ray light curves at energies above 3.5 MeV given very neutron-rich ejecta from an NSM located at 10 kpc. Line color notations are the same as Figure~\ref{fig:baselinespec}. The result given the prompt photon spectrum of $^{254}$Cf fission found when photon emission is calculated using FREYA instead of GEF (red dotted line) is also shown. 
\label{fig:baselinelc}
}
\end{figure}

To better evaluate the evolution of the MeV gamma-ray signals from fission and $\beta$-decays, we simulate the light curves above $\sim3.5$ MeV, as shown in Figure~\ref{fig:baselinelc}. Comparing the light curves before and after radiation transfer shows that for this neutron-rich outflow---and given our ejecta model and the parameter choices described in section~\ref{sec:radiationTransfer}---the ejecta becomes optically thin after $\sim$ 10 days. 
We find that, for photons in this higher MeV energy range, the gamma-ray signal from fission begins to dominate over the $\beta$-decay gamma emission after $\sim$1 day. The light-curve plateaus correspond to the fissioning species decaying on long timescales that, given the theoretical nuclear data applied in this calculation, are $^{254}$Cf between 1 day and several hundred days as well as $^{252}$Cf between 1000 and 10,000 days. Compared to the sensitivity limit of AMEGO, both the spectra and light curve suggest that a Galactic NSM would be observable by AMEGO at energies above $3.5$ MeV before $\sim$1000 days after the merger event.

We evaluate the sensitivity of our findings to the different predictions for prompt fission gamma spectra discussed in Section~\ref{sec:spectra} by examining the light curve when the FREYA calculation for $^{254}$Cf prompt gamma emission is implemented in our calculations. The gamma-ray light curve with the FREYA treatment from prompt emission has a similar shape and magnitude to that of GEF. Therefore although there remain many complexities in modeling the prompt emission from neutron-rich fission fragments, as long as the fissioning species have a nonnegligible high-energy ($> 3.5$ MeV) tail, as is supported by experiment, our conclusions regarding the the unique ability of fission to provide a signal in this energy range remain the same.

Additional nuclear physics uncertainties influence theoretical calculations of the $\beta$-decay and fission rates that determine which fissioning nuclei can be populated during the $r$ process. 
We therefore repeat our analysis using fission rates obtained with alternate fission barrier models, as in \cite{VasshGEF2019}. 
We find that although the abundance of fissioning species, as well as which fissioning nuclei play a role in late-time gamma emission, depend on the nuclear model, our main conclusions remain unchanged. 
Even the lowest predicted light curve---that obtained implementing Thomas-Fermi fission barriers, which lead to a $^{254}$Cf abundance two orders of magnitude smaller than that of Fig.~\ref{fig:baselinelc} \citep{VasshGEF2019}---is above the sensitivity limit of AMEGO from $\sim$10-100 days. We leave a more detailed investigation of the nuclear model dependence of the MeV fission gamma signature to be explored in follow-up work.

\subsection{Variations on neutron-richness and\\ the participation of fission}
\label{subsec:ye}

\begin{figure}
\includegraphics[scale=0.55]{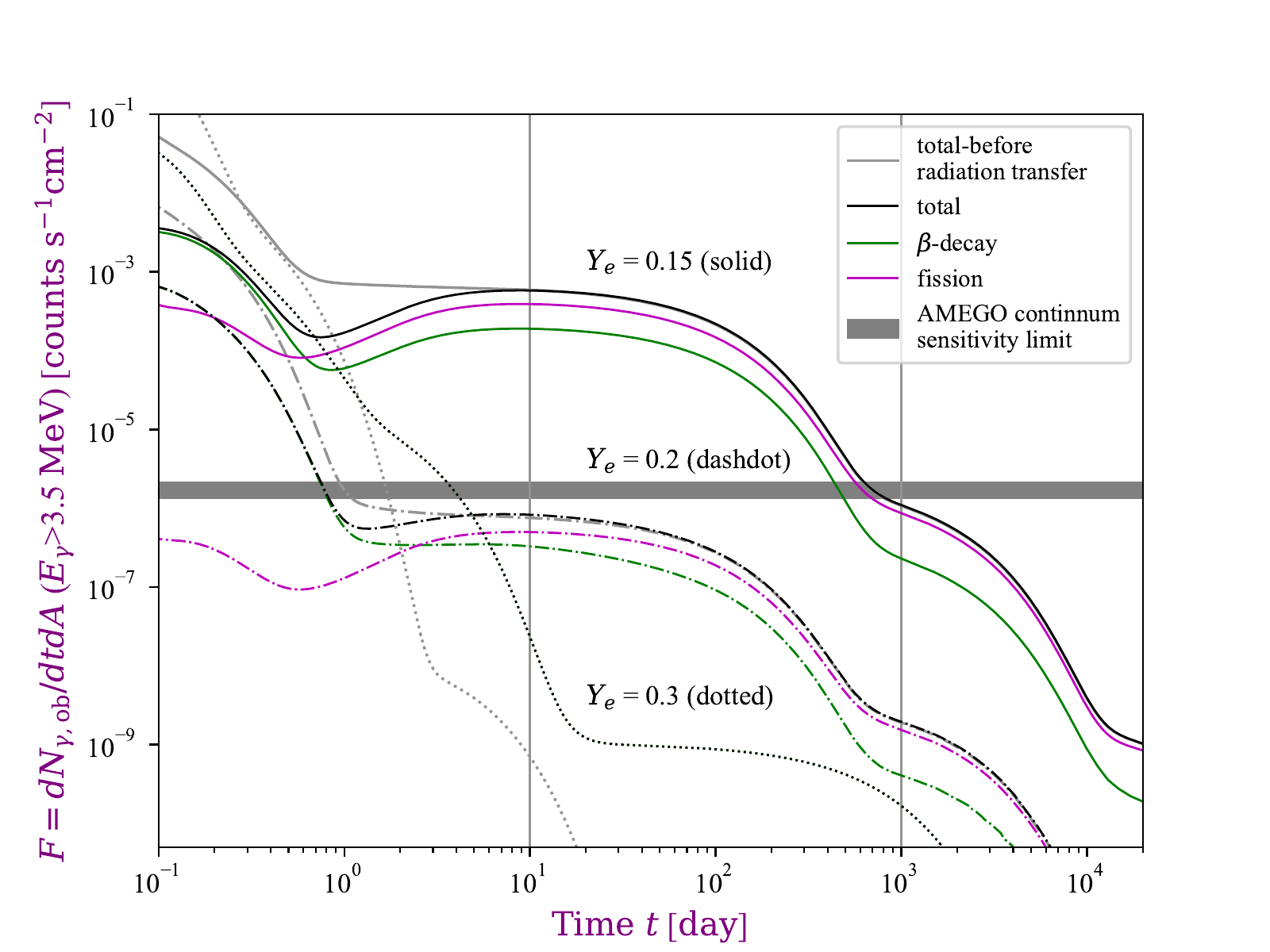}
\includegraphics[scale=0.55]{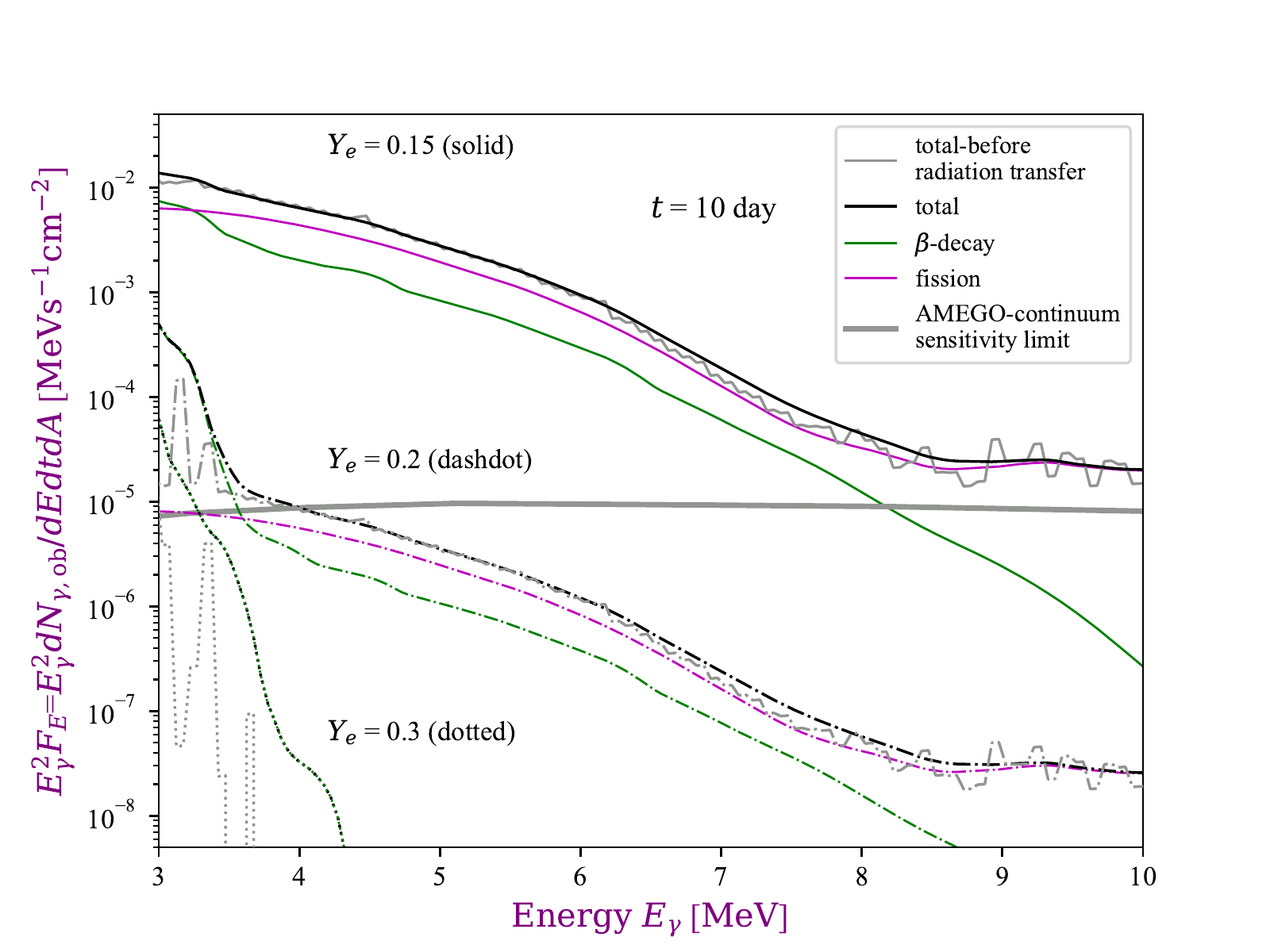}
\caption{
The predicted gamma-ray light curves at energies above 3.5 MeV (top) and the spectra at 10 days (bottom) given various levels of neutron-richness for a low entropy, long timescale outflow from an NSM located at 10 kpc. Line color notations are the same as Figure~\ref{fig:baselinespec}. An initial neutron-richness of $Y_e=0.15$ (solid line) demonstrates the case thatrobustly reaches fissioning species, whereas the $Y_e=0.2$ case (dashed-dotted line) shows a limited participation by fission and the $Y_e=0.3$ case (dotted line) shows no participation by fission.
\label{fig:winds}
}
\end{figure}

Since our results with very neutron-rich dynamical ejecta in the previous section show that outflow conditions that robustly reach fissioning species could produce a detectable MeV fission gamma-ray signal from a Galactic NSM, we next consider ejecta with a variety of neutron richness. We start with parameterized outflow conditions with low entropy ($s/k_B=30$), long timescale ($\tau=70$ ms), and a range of initial electron fractions $Y_e=0.15-0.3$, which represents conditions that could be found in both NSM dynamical and wind ejecta \citep{Just+15, Radice18}. This will produce a variation of nuclei that can be synthesized, as well as the overall abundance of actinide species, and therefore the degree to which fission participates in the $r$ process. Since the ejecta composition determines the photon opacity as well as which nuclei emit MeV gammas, we expect that MeV gamma-ray signals will be sensitive to the nature of the outflow conditions present in NSMs.

Figure~\ref{fig:winds} shows the simulated gamma-ray light curves at photon energies above $3.5$ MeV, as well as the spectra at 10 days after the merger, for outflows with $Y_e=0.15$ (robust fission), $Y_e=0.2$ (limited fission), and $Y_e=0.3$ (no fission). We see that the light curve and spectra for this type of outflow with $Y_e=0.15$ are similar to the results given the very neutron-rich ejecta considered in Figure~\ref{fig:baselinelc}, with the ejecta becoming optically thin after $\sim10$ days and the gamma-ray signal at energies above 3.5 MeV being dominated by fission over $\beta$-decay after $\sim$1 day.

In contrast, the light curves and spectra given outflow conditions that are less neutron-rich and therefore see little to no synthesis of fissioning nuclei are dramatically different from results given very neutron-rich ejecta, with the case of limited fission seeing a drop of greater than three orders of magnitude in the MeV gamma-ray signal above 3.5 MeV after $\sim$1 day.
For gammas in this energy range, the signal is strongest when conditions support a robust participation of fissioning species, and is detectable before $\sim$1000 days, whereas, in the case of limited fission, the signal is just below the detectability threshold at $\sim$10-100 days. When fissioning nuclei are not populated, the signal is too faint to be detected. Additionally, since the $\beta$ gamma spectra above $\sim 3.5$ MeV follows the fission gamma spectra in the limited fission case and becomes negligible in the case with no fission, this comparison reaffirms our previous findings that $\beta$ gamma emission at such energies is a direct consequence of fission fragments populating exotic, short-lived species at late times.

We therefore find that the late-time gamma-ray signals from NSMs in the energy range $> 3.5$ MeV are very sensitive to the degree to which fission participates in $r$-process nucleosynthesis. If such a signal is detected from a Galactic NSM at times of several days or later, the outflow conditions must be such that they support the synthesis of fissioning actinide species. Such MeV gamma-ray observations of NSMs would therefore both provide direct evidence as to whether NSMs can explain the origins of the heaviest $r$-process species, and consequently lower mass number species such as gold, as well as probe the astrophysical properties of the ejecta such as neutron richness.

\section{Discussion and Conclusions}
\label{sec: discussion}

In this paper, we report a first estimate for the contribution of fission to a late-time MeV gamma-ray signal from a Galactic NSM event. 
We find that gamma-ray emission above $3.5$ MeV at timescales longer than $\sim10$ days is a direct consequence of fissioning nuclei synthesized during the event. This signal is therefore sensitive to the fission contribution, as determined by the nature of astrophysical outflows such as neutron richness. We find that if isotopes that fission on the order of days are significantly populated, the MeV gamma-ray signal predicted is above the sensitivity limit of the AMEGO detector. Thus, by exploring the energy range distinct to fission, next generation MeV gamma-ray detectors can shed light on the synthesis achievable by NSMs. 

We note that since fission gammas in the $>3.5$ MeV energy range are primarily due to broad emission from the giant dipole resonance, we can identify no spectral lines that are unique to a particular fissioning species. Future experiments may be able to link enhanced emission lines to individual fissioning isotopes, but for now the only opportunity to identify the presence of a specific species is if it is known to have a unique half-life, as is exemplified by $^{254}$Cf and $^{252}$Cf which produce the plateaus in our light curves at timescales consistent with their half lives of 60 days and 2.6 yr respectively.

Here we have focused on fission gamma emission, which has not been considered previously in astrophysics. When viewed alongside previous studies of the lower energy signal, we find that gamma observations from a nearby NSM would  provide a wealth of information. Given the rarity of neutron star binaries in our Galaxy and the sensitivities of MeV gamma-ray telescopes, the prospects for detecting MeV gamma-ray emission from fission in NSMs are challenging. 
However, the potential pay off of such an observation is dramatic, 
and, as any Galactic event would be rarer than once-in-a-lifetime, we cannot expect a second chance.
Thus our findings call for next generation MeV detectors like AMEGO or e-ASTROGAM or LOX \citep{MeV}, to be prepared to detect the MeV signal from a Galactic NSM, which will provide a distinct and complementary approach to other observational methods aiming to understand the origin of $r$-process elements.

\acknowledgments

This work was supported by U.S. National Science Foundation under grant No. PHY-1630782 Focused Research Hub in Theoretical Physics: Network for Neutrinos, Nuclear Astrophysics, and Symmetries (N3AS) (X.W. and R.S.).
The work of N.V., M.R.M., R.V., and R.S. was partly supported by the Fission In {\em R}-process Elements (FIRE) Topical Collaboration in Nuclear Theory, funded by the U.S. Department of Energy. The work of R.V. was performed under the auspices of the U.S. Department of Energy by Lawrence Livermore National Laboratory under Contract No. DE-AC52-07NA27344.  The work of J.R. was performed under the auspices of the U.S. Department of Energy by Lawrence Berkeley National Laboratory under Contract DE-AC02-05CH11231. Additional support was provided by the U.S. Department of Energy through contract numbers DE-FG02-95-ER40934 (R.S.), and DE-SC0018232 (SciDAC TEAMS collaboration, R.S. and T.M.S). 
M.R.M was partially supported by the US Department of Energy through Los Alamos National Laboratory and by the Laboratory Directed Research and Development program of Los Alamos National Laboratory under project number 20190021DR and by the US Department of Energy through the Los Alamos National Laboratory. Los Alamos National Laboratory is operated by Triad National Security, LLC, for the National Nuclear Security Administration of U.S.\ Department of Energy (Contract No.\ 89233218CNA000001). 
T.S. was supported in part by the Los Alamos National Laboratory Center for Space and Earth Science, which is funded by its Laboratory Directed Research and Development program under project number 20180475DR. 

\bibliography{fissgamrefs}

\begin{thebibliography}{}
\expandafter\ifx\csname natexlab\endcsname\relax\def\natexlab#1{#1}\fi
\providecommand{\url}[1]{\href{#1}{#1}}
\providecommand{\dodoi}[1]{doi:~\href{http://doi.org/#1}{\nolinkurl{#1}}}
\providecommand{\doeprint}[1]{\href{http://ascl.net/#1}{\nolinkurl{http://ascl.net/#1}}}
\providecommand{\doarXiv}[1]{\href{https://arxiv.org/abs/#1}{\nolinkurl{https://arxiv.org/abs/#1}}}

\bibitem[{{Abbott} {et~al.}(2017{\natexlab{a}}){Abbott}, {Abbott}, {Abbott},
  {Acernese}, {Ackley}, {Adams}, {Adams}, {Addesso}, {Adhikari}, {Adya}, \&
  et~al.}]{NSM}
{Abbott}, B.~P., {Abbott}, R., {Abbott}, T.~D., {et~al.} 2017{\natexlab{a}},
  \apjl, 848, L12, \dodoi{10.3847/2041-8213/aa91c9}

\bibitem[{{Abbott} {et~al.}(2017{\natexlab{b}}){Abbott}, {Abbott}, {Abbott},
  {Acernese}, {Ackley}, {Adams}, {Adams}, {Addesso}, {Adhikari}, {Adya},
  {Affeldt}, {Afrough}, {Agarwal}, {Agathos}, {LIGO Scientific Collaboration},
  \& {Virgo Collaboration}}]{AbbottGW170817}
---. 2017{\natexlab{b}}, \prl, 119, 161101,
  \dodoi{10.1103/PhysRevLett.119.161101}

\bibitem[{{Billnert} {et~al.}(2013){Billnert}, {Hambsch}, {Oberstedt}, \&
  {Oberstedt}}]{Billnert}
{Billnert}, R., {Hambsch}, F.~J., {Oberstedt}, A., \& {Oberstedt}, S. 2013,
  \prc, 87, 024601, \dodoi{10.1103/PhysRevC.87.024601}

\bibitem[{{Bovard} {et~al.}(2017){Bovard}, {Martin}, {Guercilena}, {Arcones},
  {Rezzolla}, \& {Korobkin}}]{Bovard}
{Bovard}, L., {Martin}, D., {Guercilena}, F., {et~al.} 2017, \prd, 96, 124005,
  \dodoi{10.1103/PhysRevD.96.124005}

\bibitem[{Brown {et~al.}(2018)Brown, Chadwick, Capote, Kahler, Trkov, Herman,
  Sonzogni, Danon, Carlson, Dunn, Smith, Hale, Arbanas, Arcilla, Bates, Beck,
  Becker, Brown, Casperson, Conlin, Cullen, Descalle, Firestone, Gaines, Guber,
  Hawari, Holmes, Johnson, Kawano, Kiedrowski, Koning, Kopecky, Leal, Lestone,
  Lubitz, {Márquez Damián}, Mattoon, McCutchan, Mughabghab, Navratil,
  Neudecker, Nobre, Noguere, Paris, Pigni, Plompen, Pritychenko, Pronyaev,
  Roubtsov, Rochman, Romano, Schillebeeckx, Simakov, Sin, Sirakov, Sleaford,
  Sobes, Soukhovitskii, Stetcu, Talou, Thompson, {van der Marck},
  Welser-Sherrill, Wiarda, White, Wormald, Wright, Zerkle, Žerovnik, \&
  Zhu}]{ENDFB}
Brown, D., Chadwick, M., Capote, R., {et~al.} 2018, Nuclear Data Sheets, 148, 1
  , \dodoi{https://doi.org/10.1016/j.nds.2018.02.001}

\bibitem[{{Bussard} {et~al.}(1989){Bussard}, {Burrows}, \& {The}}]{Bussard1989}
{Bussard}, R.~W., {Burrows}, A., \& {The}, L.~S. 1989, \apj, 341, 401,
  \dodoi{10.1086/167503}

\bibitem[{{Churazov} {et~al.}(2014){Churazov}, {Sunyaev}, {Isern},
  {Kn{\"o}dlseder}, {Jean}, {Lebrun}, {Chugai}, {Grebenev}, {Bravo}, {Sazonov},
  \& {Renaud}}]{Churazov2014}
{Churazov}, E., {Sunyaev}, R., {Isern}, J., {et~al.} 2014, \nat, 512, 406,
  \dodoi{10.1038/nature13672}

\bibitem[{{Clayton} \& {Silk}(1969)}]{Clayton1969}
{Clayton}, D.~D., \& {Silk}, J. 1969, \apjl, 158, L43, \dodoi{10.1086/180429}

\bibitem[{{Coulter} {et~al.}(2017){Coulter}, {Foley}, {Kilpatrick}, {Drout},
  {Piro}, {Shappee}, {Siebert}, {Simon}, {Ulloa}, {Kasen}, {Madore},
  {Murguia-Berthier}, {Pan}, {Prochaska}, {Ramirez-Ruiz}, {Rest}, \&
  {Rojas-Bravo}}]{Coulter2017}
{Coulter}, D.~A., {Foley}, R.~J., {Kilpatrick}, C.~D., {et~al.} 2017, Science,
  358, 1556, \dodoi{10.1126/science.aap9811}

\bibitem[{{Cowperthwaite} {et~al.}(2017){Cowperthwaite}, {Berger}, {Villar},
  {Metzger}, {Nicholl}, {Chornock}, {Blanchard}, {Fong}, {Margutti}, \&
  {Soares-Santos}}]{Cowperthwaite2017}
{Cowperthwaite}, P.~S., {Berger}, E., {Villar}, V.~A., {et~al.} 2017, \apjl,
  848, L17, \dodoi{10.3847/2041-8213/aa8fc7}

\bibitem[{{Diehl} {et~al.}(2014){Diehl}, {Siegert}, {Hillebrandt}, {Grebenev},
  {Greiner}, {Krause}, {Kromer}, {Maeda}, {R{\"o}pke}, \&
  {Taubenberger}}]{Diehl2014}
{Diehl}, R., {Siegert}, T., {Hillebrandt}, W., {et~al.} 2014, Science, 345,
  1162, \dodoi{10.1126/science.1254738}

\bibitem[{{Drout} {et~al.}(2017){Drout}, {Piro}, {Shappee}, {Kilpatrick},
  {Simon}, {Contreras}, {Coulter}, {Foley}, {Siebert}, {Morrell}, {Boutsia},
  {Di Mille}, {Holoien}, {Kasen}, {Kollmeier}, {Madore}, {Monson},
  {Murguia-Berthier}, {Pan}, {Prochaska}, {Ramirez-Ruiz}, {Rest}, {Adams},
  {Alatalo}, {Ba{\~n}ados}, {Baughman}, {Beers}, {Bernstein}, {Bitsakis},
  {Campillay}, {Hansen}, {Higgs}, {Ji}, {Maravelias}, {Marshall}, {Moni Bidin},
  {Prieto}, {Rasmussen}, {Rojas-Bravo}, {Strom}, {Ulloa},
  {Vargas-Gonz{\'a}lez}, {Wan}, \& {Whitten}}]{Drout2017}
{Drout}, M.~R., {Piro}, A.~L., {Shappee}, B.~J., {et~al.} 2017, Science, 358,
  1570, \dodoi{10.1126/science.aaq0049}

\bibitem[{{Fryer} {et~al.}(2019){Fryer}, {Timmes}, {Hungerford}, {Couture},
  {Adams}, {Aoki}, {Arcones}, {Arnett}, {Auchettl}, {Avila}, {Badenes},
  {Baron}, {Bauswein}, {Beacom}, {Blackmon}, {Blondin}, {Bloser}, {Boggs},
  {Boss}, {Brandt}, {Bravo}, {Brown}, {Brown}, {Budtz-Jorgensen}, {Burns},
  {Calder}, {Caputo}, {Champagne}, {Chevalier}, {Chieffi}, {Chipps}, {Cinabro},
  {Clarkson}, {Clayton}, {Coc}, {Connolly}, {Conroy}, {Cote}, {Couch},
  {Dauphas}, {deBoer}, {Deibel}, {Denisenkov}, {Desch}, {Dessart}, {Diehl},
  {Doherty}, {Dominguez}, {Dong}, {Dwarkadas}, {Fan}, {Fields}, {Fields},
  {Filippenko}, {Fisher}, {Foucart}, {Fransson}, {Frohlich}, {Fuller},
  {Gibson}, {Giryanskaya}, {Gorres}, {Goriely}, {Grebenev}, {Grefenstette},
  {Grohs}, {Guillochon}, {Harpole}, {Harris}, {Harris}, {Harrison}, {Hartmann},
  {Hashimoto}, {Heger}, {Hernanz}, {Herwig}, {Hirschi}, {Hix}, {Hoflich},
  {Hoffman}, {Holcomb}, {Hsiao}, {Iliadis}, {Janiuk}, {Janka}, {Jerkstrand},
  {Johns}, {Jones}, {Jose}, {Kajino}, {Karakas}, {Karpov}, {Kasen}, {Kierans},
  {Kippen}, {Korobkin}, {Kobayashi}, {Kozma}, {Krot}, {Kumar}, {Kuvvetli},
  {Laird}, {Laming}, {Larsson}, {Lattanzio}, {Lattimer}, {Leising}, {Lennarz},
  {Lentz}, {Limongi}, {Lippuner}, {Livne}, {Lloyd-Ronning}, {Longland},
  {Lopez}, {Lugaro}, {Lutovinov}, {Madsen}, {Malone}, {Matteucci}, {McEnery},
  {Meisel}, {Messer}, {Metzger}, {Meyer}, {Meynet}, {Mezzacappa}, {Miller},
  {Miller}, {Milne}, {Misch}, {Mitchell}, {Mosta}, {Motizuki}, {Muller},
  {Mumpower}, {Murphy}, {Nagataki}, {Nakar}, {Nomoto}, {Nugent}, {Nunes},
  {O'Shea}, {Oberlack}, {Pain}, {Parker}, {Perego}, {Pignatari}, {Martinez
  Pinedo}, {Plewa}, {Poznanski}, {Priedhorsky}, {Pritychenko}, {Radice},
  {Ramirez-Ruiz}, {Rauscher}, {Reddy}, {Rehm}, {Reifarth}, {Richman}, {Ricker},
  {Rijal}, {Roberts}, {Ropke}, {Rosswog}, {Ruiter}, {Ruiz}, {Savin}, {Schatz},
  {Schneider}, {Schwab}, {Seitenzahl}, {Shen}, {Siegert}, {Sim}, {Smith},
  {Smith}, {Smith}, {Sollerman}, {Sprouse}, {Spyrou}, {Starrfield}, {Steiner},
  {Strong}, {Sukhbold}, {Suntzeff}, {Surman}, {Tanimori}, {The}, {Thielemann},
  {Tolstov}, {Tominaga}, {Tomsick}, {Townsley}, {Tsintari}, {Tsygankov},
  {Vartanyan}, {Venters}, {Vestrand}, {Vink}, {Waldman}, {Wang}, {Wang},
  {Warren}, {West}, {Wheeler}, {Wiescher}, {Winkler}, {Winter}, {Wolf},
  {Woolf}, {Woosley}, {Wu}, {Wrede}, {Yamada}, {Young}, {Zegers}, {Zingale}, \&
  {Portegies Zwart}}]{MeV}
{Fryer}, C.~L., {Timmes}, F., {Hungerford}, A.~L., {et~al.} 2019, arXiv
  e-prints, arXiv:1902.02915.
\newblock \doarXiv{1902.02915}

\bibitem[{{Goldstein} {et~al.}(2017){Goldstein}, {Veres}, {Burns}, {Briggs},
  {Hamburg}, {Kocevski}, {Wilson-Hodge}, {Preece}, {Poolakkil}, {Roberts},
  {Hui}, {Connaughton}, {Racusin}, {von Kienlin}, {Dal Canton}, {Christensen},
  {Littenberg}, {Siellez}, {Blackburn}, {Broida}, {Bissaldi}, {Cleveland},
  {Gibby}, {Giles}, {Kippen}, {McBreen}, {McEnery}, {Meegan}, {Paciesas}, \&
  {Stanbro}}]{Goldstein2017}
{Goldstein}, A., {Veres}, P., {Burns}, E., {et~al.} 2017, \apjl, 848, L14,
  \dodoi{10.3847/2041-8213/aa8f41}

\bibitem[{{Hotokezaka} {et~al.}(2016){Hotokezaka}, {Wanajo}, {Tanaka}, {Bamba},
  {Terada}, \& {Piran}}]{Hotokezaka2016}
{Hotokezaka}, K., {Wanajo}, S., {Tanaka}, M., {et~al.} 2016, \mnras, 459, 35,
  \dodoi{10.1093/mnras/stw404}

\bibitem[{{Just} {et~al.}(2015){Just}, {Bauswein}, {Pulpillo}, {Goriely}, \&
  {Janka}}]{Just+15}
{Just}, O., {Bauswein}, A., {Pulpillo}, R.~A., {Goriely}, S., \& {Janka}, H.-T.
  2015, MNRAS, 448, 541, \dodoi{10.1093/mnras/stv009}

\bibitem[{{Kasen} {et~al.}(2017){Kasen}, {Metzger}, {Barnes}, {Quataert}, \&
  {Ramirez-Ruiz}}]{Kasen}
{Kasen}, D., {Metzger}, B., {Barnes}, J., {Quataert}, E., \& {Ramirez-Ruiz}, E.
  2017, \nat, 551, 80, \dodoi{10.1038/nature24453}

\bibitem[{{Kilpatrick} {et~al.}(2017){Kilpatrick}, {Foley}, {Kasen},
  {Murguia-Berthier}, {Ramirez-Ruiz}, {Coulter}, {Drout}, {Piro}, {Shappee},
  {Boutsia}, {Contreras}, {Di Mille}, {Madore}, {Morrell}, {Pan}, {Prochaska},
  {Rest}, {Rojas-Bravo}, {Siebert}, {Simon}, \& {Ulloa}}]{Kilpatrick2017}
{Kilpatrick}, C.~D., {Foley}, R.~J., {Kasen}, D., {et~al.} 2017, Science, 358,
  1583, \dodoi{10.1126/science.aaq0073}

\bibitem[{{Korobkin} {et~al.}(2020){Korobkin}, {Hungerford}, {Fryer},
  {Mumpower}, {Misch}, {Sprouse}, {Lippuner}, {Surman}, {Couture}, {Bloser},
  {Shirazi}, {Even}, {Vestrand}, \& {Miller}}]{Oleg1}
{Korobkin}, O., {Hungerford}, A.~M., {Fryer}, C.~L., {et~al.} 2020, \apj, 889,
  168, \dodoi{10.3847/1538-4357/ab64d8}

\bibitem[{{Li}(2019)}]{Li2019}
{Li}, L.-X. 2019, \apj, 872, 19, \dodoi{10.3847/1538-4357/aaf961}

\bibitem[{{Matz} {et~al.}(1988){Matz}, {Share}, {Leising}, {Chupp}, \&
  {Vestrand}}]{Matz1988}
{Matz}, S.~M., {Share}, G.~H., {Leising}, M.~D., {Chupp}, E.~L., \& {Vestrand},
  W.~T. 1988, \nat, 331, 416, \dodoi{10.1038/331416a0}

\bibitem[{{McEnery} {et~al.}(2019){McEnery}, {van der Horst}, {Dominguez},
  {Moiseev}, {Marcowith}, {Harding}, {Lien}, {Giuliani}, {Inglis}, {Ansoldi},
  {Stamerra}, {Manousakis}, {Strong}, {Bambi}, {Patricelli}, {Baring},
  {Barrio}, {Bastieri}, {Fields}, {Beacom}, {Beckmann}, {Bednarek}, {Rani},
  {Boggs}, {Bolotnikov}, {Cenko}, {Buckley}, {Grefenstette}, {Hui}, {Pittori},
  {Prescod-Weinstein}, {Shrader}, {Gouiffes}, {Kierans}, {Wilson-Hodge},
  {D'Ammando}, {Castro}, {Kocveski}, {Gasparrini}, {Thompson}, {Williams}, {De
  Angelis}, {Bernard}, {Digel}, {Morcuende}, {Charles}, {Bissaldi}, {Hays},
  {Ferrara}, {Bozzo}, {Grove}, {Wulf}, {Bottacini}, {Caroli}, {Kislat},
  {Oikonomou}, {Giordano}, {Longo}, {Fryer}, {Fukazawa}, {Georganopoulos}, {De
  Nolfo}, {Vianello}, {Kanbach}, {Younes}, {Blumer}, {Hartmann}, {Hernanz},
  {Takahashi}, {Li}, {Agudo}, {Moskalenko}, {Stumke}, {Grenier}, {Smith},
  {Rodi}, {Perkins}, {Gelfand}, {Holder}, {Knodlseder}, {Kopp}, {Lenain},
  {{\'A}lvarez}, {Metcalfe}, {Krizmanic}, {Stephen}, {Hewitt}, {Mitchell},
  {Harding}, {Tomsick}, {Racusin}, {Finke}, {Kargaltsev}, {Klimenko},
  {Krawczynski}, {Smith}, {Kubo}, {Di Venere}, {Marcotulli}, {Lommler},
  {Parker}, {Baldini}, {Foffano}, {Zampieri}, {Tibaldo}, {Petropoulou},
  {Ajello}, {Meyer}, {L{\'o}pez}, {McConnell}, {Boettcher}, {Cardillo},
  {Martinez}, {Kerr}, {Mazziotta}, {McEnery}, {Di Mauro}, {Wood}, {Meyer},
  {Briggs}, {De Becker}, {Lovellette}, {Doro}, {Sanchez-Conde}, {Moss},
  {Mizuno}, {Rib{\'o}}, {Nakazawa}, {Neilson}, {Auricchio}, {Omodei},
  {Oberlack}, {Ohno}, {Orland o}, {Otte}, {Coppi}, {Bloser}, {Zhang},
  {Laurent}, {Pohl}, {Prand ini}, {Shawhan}, {Caputo}, {Campana}, {Rando},
  {Woolf}, {Johnson}, {Mignani}, {Walter}, {Ojha}, {da Silva}, {Dietrich},
  {Funk}, {Zane}, {Anton}, {Buson}, {Cutini}, {Saz Parkinson}, {Schirato},
  {Griffin}, {Kaufmann}, {Stawarz}, {Ciprini}, {Del Sordo}, {Jones}, {Guiriec},
  {Tajima}, {Cheung}, {The}, {Venters}, {Porter}, {Linden}, {Barres}, {Paliya},
  {Bozhilov}, {Vestrand}, {Tatischeff}, {Chen}, {Wang}, {Tanaka}, {Uhm},
  {Zhang}, {Zimmer}, {Zoglauer}, \& {Wadiasingh}}]{AMEGO}
{McEnery}, J., {van der Horst}, A., {Dominguez}, A., {et~al.} 2019, in Bulletin
  of the American Astronomical Society, Vol.~51, 245

\bibitem[{{M{\"o}ller} {et~al.}(2019){M{\"o}ller}, {Mumpower}, {Kawano}, \&
  {Myers}}]{MollerQRPA}
{M{\"o}ller}, P., {Mumpower}, M.~R., {Kawano}, T., \& {Myers}, W.~D. 2019, At.\
  Data Nucl.\ Data Tables, 125, 1,
  \dodoi{https://doi.org/10.1016/j.adt.2018.03.003}

\bibitem[{{Mumpower} {et~al.}(2016){Mumpower}, {Kawano}, \&
  {M{\"o}ller}}]{Mumpower+16}
{Mumpower}, M.~R., {Kawano}, T., \& {M{\"o}ller}, P. 2016, \prc, 94, 064317,
  \dodoi{10.1103/PhysRevC.94.064317}

\bibitem[{{Mumpower} {et~al.}(2018){Mumpower}, {Kawano}, {Sprouse}, {Vassh},
  {Holmbeck}, {Surman}, \& {M{\"o}ller}}]{BDFrp}
{Mumpower}, M.~R., {Kawano}, T., {Sprouse}, T.~M., {et~al.} 2018, ApJ, 869, 14,
  \dodoi{10.3847/1538-4357/aaeaca}

\bibitem[{{Nomoto} {et~al.}(2013){Nomoto}, {Kobayashi}, \&
  {Tominaga}}]{Nomoto2013}
{Nomoto}, K., {Kobayashi}, C., \& {Tominaga}, N. 2013, \araa, 51, 457,
  \dodoi{10.1146/annurev-astro-082812-140956}

\bibitem[{{Nomoto} {et~al.}(1984){Nomoto}, {Thielemann}, \&
  {Yokoi}}]{Nomoto1984}
{Nomoto}, K., {Thielemann}, F.~K., \& {Yokoi}, K. 1984, \apj, 286, 644,
  \dodoi{10.1086/162639}

\bibitem[{{Pian} {et~al.}(2017){Pian}, {D'Avanzo}, {Benetti}, {Branchesi},
  {Brocato}, {Campana}, {Cappellaro}, {Covino}, {D'Elia}, {Fynbo}, {Getman},
  {Ghirland a}, {Ghisellini}, {Grado}, {Greco}, {Hjorth}, {Kouveliotou},
  {Levan}, {Limatola}, {Malesani}, {Mazzali}, {Melandri}, {M{\o}ller},
  {Nicastro}, {Palazzi}, {Piranomonte}, {Rossi}, {Salafia}, {Selsing},
  {Stratta}, {Tanaka}, {Tanvir}, {Tomasella}, {Watson}, {Yang}, {Amati},
  {Antonelli}, {Ascenzi}, {Bernardini}, {Bo{\"e}r}, {Bufano}, {Bulgarelli},
  {Capaccioli}, {Casella}, {Castro-Tirado}, {Chassande-Mottin}, {Ciolfi},
  {Copperwheat}, {Dadina}, {De Cesare}, {di Paola}, {Fan}, {Gendre},
  {Giuffrida}, {Giunta}, {Hunt}, {Israel}, {Jin}, {Kasliwal}, {Klose}, {Lisi},
  {Longo}, {Maiorano}, {Mapelli}, {Masetti}, {Nava}, {Patricelli}, {Perley},
  {Pescalli}, {Piran}, {Possenti}, {Pulone}, {Razzano}, {Salvaterra},
  {Schipani}, {Spera}, {Stamerra}, {Stella}, {Tagliaferri}, {Testa}, {Troja},
  {Turatto}, {Vergani}, \& {Vergani}}]{Pian2017}
{Pian}, E., {D'Avanzo}, P., {Benetti}, S., {et~al.} 2017, \nat, 551, 67,
  \dodoi{10.1038/nature24298}

\bibitem[{{Piran} {et~al.}(2013){Piran}, {Nakar}, \& {Rosswog}}]{PiranRoss}
{Piran}, T., {Nakar}, E., \& {Rosswog}, S. 2013, Mon.\ Not.\ R.\ Astron.\ Soc.,
  430, 2121, \dodoi{10.1093/mnras/stt037}

\bibitem[{{Qi} {et~al.}(2018){Qi}, {Lebois}, {Wilson}, {Chatillon}, {Courtin},
  {Fruet}, {Georgiev}, {Jenkins}, {Laurent}, {Le Meur}, {Maj}, {Marini},
  {Matea}, {Morris}, {Nanal}, {Napiorkowski}, {Oberstedt}, {Oberstedt},
  {Schmitt}, {Serot}, {Stanoiu}, \& {Wasilewska}}]{Qi}
{Qi}, L., {Lebois}, M., {Wilson}, J.~N., {et~al.} 2018, \prc, 98, 014612,
  \dodoi{10.1103/PhysRevC.98.014612}

\bibitem[{{Radice} {et~al.}(2018){Radice}, {Perego}, {Hotokezaka}, {Fromm},
  {Bernuzzi}, \& {Roberts}}]{Radice18}
{Radice}, D., {Perego}, A., {Hotokezaka}, K., {et~al.} 2018, \apj, 869, 130,
  \dodoi{10.3847/1538-4357/aaf054}

\bibitem[{{Rosswog} {et~al.}(2013){Rosswog}, {Piran}, \& {Nakar}}]{Rosswog}
{Rosswog}, S., {Piran}, T., \& {Nakar}, E. 2013, Mon.\ Not.\ R.\ Astron.\ Soc.,
  430, 2585, \dodoi{10.1093/mnras/sts708}

\bibitem[{{Ruiz-Lapuente} \& {Korobkin}(2020)}]{Oleg2}
{Ruiz-Lapuente}, P., \& {Korobkin}, O. 2020, \apj, 892, 45,
  \dodoi{10.3847/1538-4357/ab744e}

\bibitem[{{Savchenko} {et~al.}(2017){Savchenko}, {Ferrigno}, {Kuulkers},
  {Bazzano}, {Bozzo}, {Brandt}, {Chenevez}, {Courvoisier}, {Diehl}, {Domingo},
  {Hanlon}, {Jourdain}, {von Kienlin}, {Laurent}, {Lebrun}, {Lutovinov},
  {Martin-Carrillo}, {Mereghetti}, {Natalucci}, {Rodi}, {Roques}, {Sunyaev}, \&
  {Ubertini}}]{Savchenko2017}
{Savchenko}, V., {Ferrigno}, C., {Kuulkers}, E., {et~al.} 2017, \apjl, 848,
  L15, \dodoi{10.3847/2041-8213/aa8f94}

\bibitem[{{Schmidt} {et~al.}(2016){Schmidt}, {Jurado}, {Amouroux}, \&
  {Schmitt}}]{GEF}
{Schmidt}, K.-H., {Jurado}, B., {Amouroux}, C., \& {Schmitt}, C. 2016, Nucl.\
  Data Sheets, 131, 107, \dodoi{10.1016/j.nds.2015.12.009}

\bibitem[{{Shappee} {et~al.}(2017){Shappee}, {Simon}, {Drout}, {Piro},
  {Morrell}, {Prieto}, {Kasen}, {Holoien}, {Kollmeier}, {Kelson}, {Coulter},
  {Foley}, {Kilpatrick}, {Siebert}, {Madore}, {Murguia-Berthier}, {Pan},
  {Prochaska}, {Ramirez-Ruiz}, {Rest}, {Adams}, {Alatalo}, {Ba{\~n}ados},
  {Baughman}, {Bernstein}, {Bitsakis}, {Boutsia}, {Bravo}, {Di Mille}, {Higgs},
  {Ji}, {Maravelias}, {Marshall}, {Placco}, {Prieto}, \& {Wan}}]{Shappee2017}
{Shappee}, B.~J., {Simon}, J.~D., {Drout}, M.~R., {et~al.} 2017, Science, 358,
  1574, \dodoi{10.1126/science.aaq0186}

\bibitem[{{Teegarden} {et~al.}(1989){Teegarden}, {Barthelmy}, {Gehrels},
  {Tueller}, \& {Leventhal}}]{Teegarden1989}
{Teegarden}, B.~J., {Barthelmy}, S.~D., {Gehrels}, N., {Tueller}, J., \&
  {Leventhal}, M. 1989, \nat, 339, 122, \dodoi{10.1038/339122a0}

\bibitem[{{The} \& {Burrows}(2014)}]{The2014}
{The}, L.-S., \& {Burrows}, A. 2014, \apj, 786, 141,
  \dodoi{10.1088/0004-637X/786/2/141}

\bibitem[{{Thielemann} {et~al.}(2017){Thielemann}, {Eichler}, {Panov}, \&
  {Wehmeyer}}]{Thielemann2017}
{Thielemann}, F.~K., {Eichler}, M., {Panov}, I.~V., \& {Wehmeyer}, B. 2017,
  Annual Review of Nuclear and Particle Science, 67, 253,
  \dodoi{10.1146/annurev-nucl-101916-123246}

\bibitem[{{Valenti} {et~al.}(2017){Valenti}, {Sand}, {Yang}, {Cappellaro},
  {Tartaglia}, {Corsi}, {Jha}, {Reichart}, {Haislip}, \&
  {Kouprianov}}]{Valenti2017}
{Valenti}, S., {Sand}, D.~J., {Yang}, S., {et~al.} 2017, \apjl, 848, L24,
  \dodoi{10.3847/2041-8213/aa8edf}

\bibitem[{{Vassh} {et~al.}(2019){Vassh}, {Vogt}, {Surman}, {Randrup},
  {Sprouse}, {Mumpower}, {Jaffke}, {Shaw}, {Holmbeck}, {Zhu}, \&
  {McLaughlin}}]{VasshGEF2019}
{Vassh}, N., {Vogt}, R., {Surman}, R., {et~al.} 2019, Journal of Physics G
  Nuclear Physics, 46, 065202, \dodoi{10.1088/1361-6471/ab0bea}

\bibitem[{{Verbeke} {et~al.}(2018){Verbeke}, {Randrup}, \& {Vogt}}]{FREYA2}
{Verbeke}, J.~M., {Randrup}, J., \& {Vogt}, R. 2018, Comp.\ Phys.\ Comm., 222,
  263, \dodoi{10.1016/j.cpc.2017.09.006}

\bibitem[{Vogt \& Randrup(2017)}]{RVJR_gamma2}
Vogt, R., \& Randrup, J. 2017, \prc, 96, 064620,
  \dodoi{10.1103/PhysRevC.96.064620}

\bibitem[{{Wang} {et~al.}(2019){Wang}, {Fields}, \& {Lien}}]{Wang2019}
{Wang}, X., {Fields}, B.~D., \& {Lien}, A.~Y. 2019, \mnras, 486, 2910,
  \dodoi{10.1093/mnras/stz993}

\bibitem[{{Wang} {et~al.}(2020){Wang}, {Fields}, {Mumpower}, {Sprouse},
  {Surman}, \& {Vassh}}]{Wang2020}
{Wang}, X., {Fields}, B.~D., {Mumpower}, M., {et~al.} 2020, \apj, 893, 92,
  \dodoi{10.3847/1538-4357/ab7ffd}

\bibitem[{{Watson} {et~al.}(2019){Watson}, {Hansen}, {Selsing}, {Koch},
  {Malesani}, {Andersen}, {Fynbo}, {Arcones}, {Bauswein}, {Covino}, {Grado},
  {Heintz}, {Hunt}, {Kouveliotou}, {Leloudas}, {Levan}, {Mazzali}, \&
  {Pian}}]{Watson2019}
{Watson}, D., {Hansen}, C.~J., {Selsing}, J., {et~al.} 2019, \nat, 574, 497,
  \dodoi{10.1038/s41586-019-1676-3}

\bibitem[{{Waxman} {et~al.}(2019){Waxman}, {Ofek}, \& {Kushnir}}]{Waxman2019}
{Waxman}, E., {Ofek}, E.~O., \& {Kushnir}, D. 2019, \apj, 878, 93,
  \dodoi{10.3847/1538-4357/ab1f71}

\bibitem[{{Wu} {et~al.}(2019{\natexlab{a}}){Wu}, {Barnes},
  {Mart{\'\i}nez-Pinedo}, \& {Metzger}}]{Wu1}
{Wu}, M.-R., {Barnes}, J., {Mart{\'\i}nez-Pinedo}, G., \& {Metzger}, B.~D.
  2019{\natexlab{a}}, \prl, 122, 062701, \dodoi{10.1103/PhysRevLett.122.062701}

\bibitem[{{Wu} {et~al.}(2019{\natexlab{b}}){Wu}, {Banerjee}, {Metzger},
  {Mart{\'\i}nez-Pinedo}, {Aramaki}, {Burns}, {Hailey}, {Barnes}, \&
  {Karagiorgi}}]{Wu2}
{Wu}, M.-R., {Banerjee}, P., {Metzger}, B.~D., {et~al.} 2019{\natexlab{b}},
  \apj, 880, 23, \dodoi{10.3847/1538-4357/ab2593}

\bibitem[{{Zhu} {et~al.}(2018){Zhu}, {Wollaeger}, {Vassh}, {Surman}, {Sprouse},
  {Mumpower}, {M{\"o}ller}, {McLaughlin}, {Korobkin}, {Kawano}, {Jaffke},
  {Holmbeck}, {Fryer}, {Even}, {Couture}, \& {Barnes}}]{Cfpaper}
{Zhu}, Y., {Wollaeger}, R.~T., {Vassh}, N., {et~al.} 2018, ApJL, 863, L23,
  \dodoi{10.3847/2041-8213/aad5de}

\end{thebibliography}

\end{document}